\documentclass[aps,superscriptaddress,amsmath,pra,amssymb,raggedfooter,reprint]{revtex4-2}

\usepackage[USenglish]{babel}
\usepackage[T1]{fontenc}
\usepackage[utf8]{inputenc}
\usepackage{graphicx}
\usepackage{upgreek}
\usepackage[unicode]{hyperref}
\begin{document}
	
\title{Suppression of nuclear spin fluctuations in an InGaAs quantum dot ensemble by GHz-pulsed optical excitation}
		
\date{\today}
	
\author{E.~Evers}
\email[Email: ]{eiko.evers@tu-dortmund.de}
\affiliation{Experimentelle Physik 2, Technische Universit\"at Dortmund, 44221 Dortmund, Germany}
	
\author{N.~E.~Kopteva}
\affiliation{Experimentelle Physik 2, Technische Universit\"at Dortmund, 44221 Dortmund, Germany}

\author{I.~A.~Yugova}
\affiliation{Spin Optics Laboratory, St. Petersburg State University, 198504 St.\,Petersburg, Russia}

\author{D.~R.~Yakovlev}
\affiliation{Experimentelle Physik 2, Technische Universit\"at Dortmund, 44221 Dortmund, Germany}
\affiliation{Ioffe Institute, Russian Academy of Sciences, 194021 St.\,Petersburg, Russia}

\author{D.~Reuter}
\altaffiliation[Present address: ]{Department Physik, Universität Paderborn, 33098 Paderborn, Germany}
\affiliation{Angewandte Festkörperphysik, Ruhr-Universität Bochum, 44780 Bochum, Germany}

\author{A.~D.~Wieck}
\affiliation{Angewandte Festkörperphysik, Ruhr-Universität Bochum, 44780 Bochum, Germany}

\author{M.~Bayer}
\affiliation{Experimentelle Physik 2, Technische Universit\"at Dortmund, 44221 Dortmund, Germany}
\affiliation{Ioffe Institute, Russian Academy of Sciences, 194021 St.\,Petersburg, Russia}

\author{A.~Greilich}
\affiliation{Experimentelle Physik 2, Technische Universit\"at Dortmund, 44221 Dortmund, Germany}

\begin{abstract}
The coherent electron spin dynamics of an ensemble of singly charged (In,Ga)As/GaAs quantum dots in a transverse magnetic field is driven by periodic optical excitation at 1\,GHz repetition frequency. Despite the strong inhomogeneity of the electron $g$ factor, the spectral spread of optical transitions, and the broad distribution of nuclear spin fluctuations, we are able to push the whole ensemble of excited spins into a single Larmor precession mode that is commensurate with the laser repetition frequency. Furthermore, we demonstrate that an optical detuning of the pump pulses from the probed optical transitions induces a directed dynamic nuclear polarization and leads to a discretization of the total magnetic field acting on the electron ensemble. Finally, we show that the highly periodic optical excitation can be used as universal tool for strongly reducing the nuclear spin fluctuations and preparation of a robust nuclear environment for subsequent manipulation of the electron spins, also at varying operation frequencies.
\end{abstract}

\maketitle
\section*{Introduction}
The last decade has been marked by unprecedented progress in the development of quantum technologies. This is confirmed by the development and first implementation of quantum communication~\cite{Satellite17} and quantum computing~\cite{Arute2019}. At the heart of these technologies are solid state quantum bits (qubits) and their entanglement~\cite{Ladd2010}. As the race for the best qubit candidate is still ongoing, it becomes clear that there will be no monolithic solution, but rather a hybrid solution combining different excitations, each exploiting its own best property while contributing to the common goal of the targeted quantum technology.

One of the possible hybrid qubit realizations is the spin of an electron confined in a semiconductor quantum dot (QD), which is interacting with the surrounding nuclear spins~\cite{Ladd2010}. The prominent advantage of QDs is their strong optical dipole moment, which allows efficient coupling of photons to the confined electron spins, according to optical selection rules. The electron spin is coupled to the nuclear spins of the QD crystal lattice by the hyperfine interaction~\cite{MerkulovEfrosRosen}, which could allow one to design schemes where the angular momentum of the photon is transferred to the nuclear spins using the electron spin as auxiliary state. The advantage of this approach is that the electron spin coherence is limited to several microseconds at low temperatures~\cite{Greilich2006a}, but the nuclear spin coherence can last milliseconds~\cite{Wuest2016}, allowing in particular the implementation of quantum repeater schemes~\cite{Wehnereaam9288}.

The idea to transfer the electron spin state to the surrounding nuclear spins is aggravated by the intrinsic nuclear spin fluctuations~\cite{MerkulovEfrosRosen}. A way to reduce these fluctuations was first elaborated theoretically~\cite{StepanenkoPRL06} and later demonstrated in a series of experiments~\cite{XuNature09,LattaNatPhys09,OnurPRB16,EthierPRL17,Bodey2019}. Further advancement in the reduction of nuclear spin fluctuations led to the possibility to implement all-optical access to the individual quantized transitions of the strongly coupled electron-nuclear spin systems~\cite{GangloffScience19}. All these experiments were realized on single QDs and required a high spectral precision. 

In this paper, we explore an alternative and universal tool that has relaxed requirements on spectral and other material contents-related differences of single QDs. Using a single pulsed laser source it becomes possible to control the state of all QDs whose optical transitions fall into the spectrum of the laser at the same time. To prove its universality, we apply our method to an ensemble of QDs and detect their joint response. We expose this inhomogeneous ensemble of singly-charged (In,Ga)As/GaAs QDs to a high repetition laser operated at 1\,GHz rate. Exploiting the strong electron-nuclear feedback we drive the inhomogeneous ensemble of electron spins into single frequency Larmor precession about a transverse magnetic field. Additionally, we demonstrate the discretization of the total magnetic field acting on the electron spin ensemble and demonstrate a reduction of the nuclear spin fluctuations, which leads to a deceleration of the electron spin dephasing. We further demonstrate that one can prepare the QD system using a high repetition rate excitation in such a low dephasing-state and then switch non-detrimentally to another laser source operating at a different repetition frequency for subsequent manipulation. This can be done on time scales up to seconds to continue manipulation of the electron spins in the reduced fluctuation environment.

\section*{Results}
\subsection*{Prolongation of the spin dephasing time $T_2^*$}
\begin{figure*}[t]
\begin{center}
\includegraphics[trim=0mm 0mm 0mm 0mm, clip, width=2.05\columnwidth]{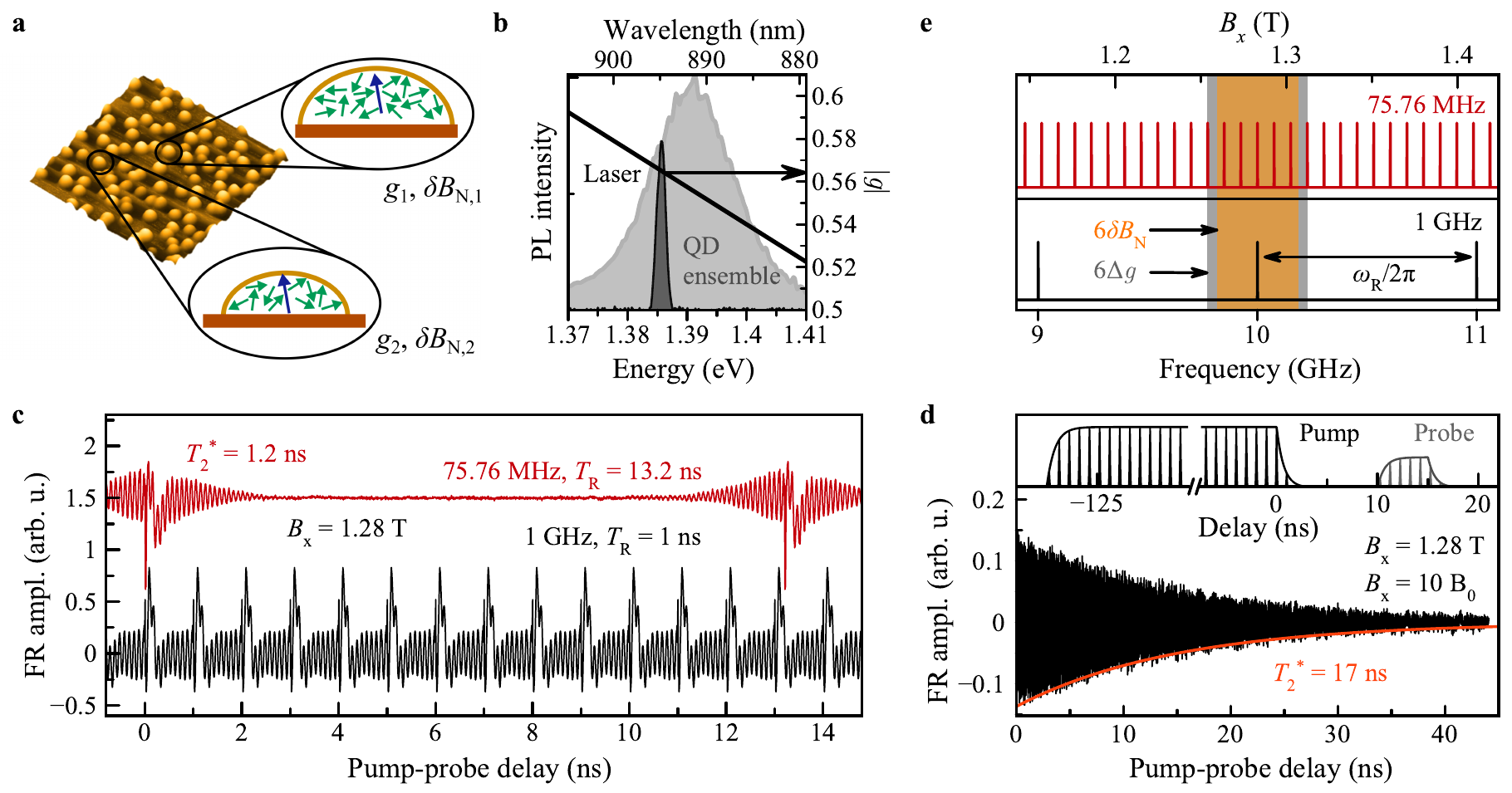}
\caption{\textbf{Dependence of spin dephasing time $T_2^*$ on laser repetition frequency.} \textbf{a} Atomic force microscope of the (In,Ga)As/GaAs QD ensemble with a QD density of $10^{10}$\,cm$^{-2}$. Sketches indicate the nuclear spins (green arrows) interacting with a single electron (blue arrow). The QDs differ in size as well as in $g$-factor value. \textbf{b} Photoluminescence~(PL) spectrum of the (In,Ga)As/GaAs QD ensemble (light gray) and spectrum of the picosecond laser emission, shown by the dark gray shaded profile. The black line gives the dispersion of the absolute electron $g$ factor. \textbf{c} Time-resolved Faraday rotation for laser repetition frequencies of 75.76$\,$MHz (red) and 1$\,$GHz (black). \textbf{d} Extended pump-probe signal with $T_2^*=17$\,ns. The inset shows the sequences of the applied pump and probe pulse bunches. For panels (\textbf{c}, \textbf{d}) containing experimental data the following conditions hold: $T=5.3$\,K, $E_{\text{Pu/Pr}}=1.3867$\,eV. Used laser powers: for 75.76\,MHz $P_\text{Pu}=600$\,W/cm$^{2}$, $P_\text{Pr}=6$\,W/cm$^{2}$; for 1\,GHz $P_\text{Pu}=260$\,W/cm$^{2}$, $P_\text{Pr}=8$\,W/cm$^{2}$. \textbf{e} Scheme of the PSC fulfilling precession frequencies (red lines) for the 75.76\,MHz laser. The orange area shows the spread of the nuclear fluctuations in the QDs, $6\delta B_{\text{N}}=45$\,mT, and the gray shaded area gives the precession frequency range due to $g$-factor spread at $B_x=1.28$\,T. The black lines below demonstrate the situation for the 1\,GHz laser excitation. Here, the mode separation is increased to 128\,mT.}
\label{fig1}
\end{center}
\end{figure*}

An ensemble of self-assembled QDs is known to be intrinsically inhomogeneous. Figure~\ref{fig1}a shows an atomic force microscopy image of the studied QD ensemble, from which the variation of the QD sizes becomes apparent. This inhomogeneity leads to the broad emission spectrum shown by the light gray shaded trace in Fig.~\ref{fig1}b. Furthermore, every QD in the ensemble contains about $10^5$ nuclear spins, so that one expects nuclear-spin fluctuations in the Overhauser field ($\delta B_{\text{N}}$)~\cite{MerkulovEfrosRosen,Urbaszek2013} acting on the electron spins in the QDs due to the hyperfine interaction, see sketches in Fig.~\ref{fig1}a. Due to the variation of the constituent material, the inhomogeneity is also present in the electron $g$ factors in the ensemble, whose dispersion is shown by the black line in Fig.~\ref{fig1}b. The combination of these effects manifests itself as a fast dephasing of the measured ensemble spin dynamics in magnetic field, occuring on the timescale $T_2^*$ of a nanosecond~\cite{Greilich2006a,BechtoldNatPhys15}. These inhomogneities can, however, be overcome by the experiment's design, exploiting the effects of spin mode-locking (SML)~\cite{Greilich2006a} and nuclear induced frequency focusing (NIFF)~\cite{Greilich2007,MarkmannNatCom19}.

To study the coherent spin dynamics in the QD ensemble, we use time-resolved Faraday rotation (FR). Exemplary traces for pulsed excitation with repetition frequencies of $75.76\,$MHz (red) and $1\,$GHz (black), corresponding to repetition periods of $T_{\text{R}} = 13.2\,$ns and $T_{\text{R}} = 1\,$ns, respectively, are shown in Fig.~\ref{fig1}c for $B_x = 1.28\,$T. As one can see, for the case of $T_{\text{R}} = 13.2\,$ns the signal decays within $T_2^*= 1.2\,$ns, while there is no observable spin decay for 1\,ns pulse separation. Here, an assessment of the temporal dynamics is impossible for times exceeding 1\,ns, therefore we apply an adapted extended pump-probe method~\cite{Belykh2016}. The spin dynamics is shown in Fig.~\ref{fig1}d, demonstrating electron spin dephasing on a timescale of $T_2^*=17\,$ns. In this case, pump and probe pulses are picked by electro-optical modulators and hit the sample in bunches, with a controlled delay time between the pump and probe pulse combinations, see the inset in Fig.~\ref{fig1}d. Additional data can be found in the Supplementary Note 1.

To explain the observed difference in $T_2^*$ for both repetition frequencies, we first consider the case of $13.2\,$ns repetition period. The FR signal exhibits a pronounced rise of the electron spin polarization before each pump pulse arrival ($0\,$ns or $13.2\,$ns delay) which mirrors the decay thereafter (effect of SML)~\cite{Greilich2006a}. Both the decay and the rise of the signal are caused by the superposition of multiple precession modes which leads to destructive signal interference between the pump pulses. At a delay of a multiple integer of $T_{\text{R}}$, constructive interference occurs for particular modes with discrete electron spin precession frequencies $\omega$. Generally, $\omega=g \mu_{\rm B} B_x/\hbar$ in the external magnetic field $B_x$, where $\mu_{\rm B}$ is the Bohr magneton and $\hbar$ is the reduced Planck constant. The frequencies of the constructively interfering precession modes satisfy the phase synchronization condition (PSC) $\omega = K\omega_{\text{R}}$, where $\omega_{\text{R}} = 2\pi/T_{\text{R}}$ is the repetition rate of the laser pulses and $K$ is an integer characterizing each contributing mode. As discussed in Refs.~\cite{Greilich2009a,PSS09}, the number of PSC precession modes, $M$, within the inhomogeneous ensemble is given by: (1) the $g$-factor spread of the optically excited electron spins ($\Delta g$), (2) the nuclear spin fluctuations ($\delta B_\text{N}$), (3) the external magnetic field ($B_x$), and (4) the laser repetition period $T_{\text{R}}$. 

The black solid line in Fig.~\ref{fig1}b demonstrates the dependence of the electron $g$ factor on the optical excitation energy, following roughly a linear dependence with a slope of $\Delta g/\Delta E = -1.75$\,eV$^{-1}$~\cite{Fine_structure,Yugova2009}. Using the laser energy and the spectral pulse width, this dependence allows us to determine the average $g$ factor at the probe energy of $1.3867\,$eV to be $|g|=0.57$ with a spread of $\Delta g=0.004$~\cite{Greilich2006}. The nuclear field fluctuations are known to be $\delta B_{\text{N}}=7.5\,$mT for this sample~\cite{Greilich2009a}. Therefore, the number of contributing PSC modes at $B_x=1.28\,$T is dominated by the $g$-factor spread which covers $M=8$ modes for $T_{\text{R}}=13.2\,$ns, as shown by the gray-shaded area in Fig.~\ref{fig1}e~\cite{Evers18}. 

The number of modes $M$ is derived here for a width of the Gaussian precession frequency distribution taken as six times its half width at half maximum (HWHM), to account for 99.7\,\% of the spins~\cite{Evers18}:
\begin{equation}
M=6\Delta \omega T_{\text{R}} / 2 \pi,
\label{eqM}
\end{equation}
where
\begin{equation}
\Delta \omega = \mu_B\sqrt{(\Delta g B_x)^2+(g \delta B_{\text{N}})^2}/\hbar.
\label{eqM2}
\end{equation}

The nuclear fluctuation field $\delta B_{\text{N}}$ shown by the orange-shaded area is dominant only at small magnetic fields~\cite{Greilich2009a}, where $\Delta g$ does not contribute significantly anymore. For $T_{\text{R}}=13.2\,$ns, the number of modes covered by the frequency distribution is larger than unity for any magnetic field strength, causes generally fast spin dephasing.

The situation for 1\,GHz laser repetition rate is shown by the black trace in Fig.~\ref{fig1}c. For pulsed excitation with $T_{\text{R}} = 1\,$ns, the separation between neighboring PSC modes is $B_0 = \hbar \omega_{\text{R}} /(g \mu_{\rm B}) = 128$\,mT, which is much larger than the $\delta B_\text{N}$ of the nuclear spin fluctuations (7.5\,mT). Moreover, the $g$-factor spread is also not sufficient to allow for more than one mode within the $128\,$mT range at a field of $B_x = 1.28\,$T (see the black lines in Fig.~\ref{fig1}e). As a result, the signal shows a single, slowly decaying oscillation with $T_2^*=17\,$ns instead of a multi-mode signal with fast dephasing of $1\,$ns. Hence, the pump-probe signal between two pump pulses for 1\,GHz excitation can be evaluated using a single cosine function with a frequency $\omega$:
\begin{equation}
S(t) = S_1\cos(\omega t).
\label{eq1}
\end{equation}
$S$ is the signal amplitude, $S_1 = S_0\exp{(-t/T_2^*)}$ where $S_0$ is the electron spin polarization created by the pump, $t$ is the pump-probe time delay and $T_2^*$ is the electron spin dephasing time related to the single-mode frequency bandwidth.

\subsection*{Influence of nuclear spins}
\begin{figure*}[t]
	\begin{center}
		\includegraphics[trim=0mm 0mm 0mm 0mm, clip, width=2.05\columnwidth]{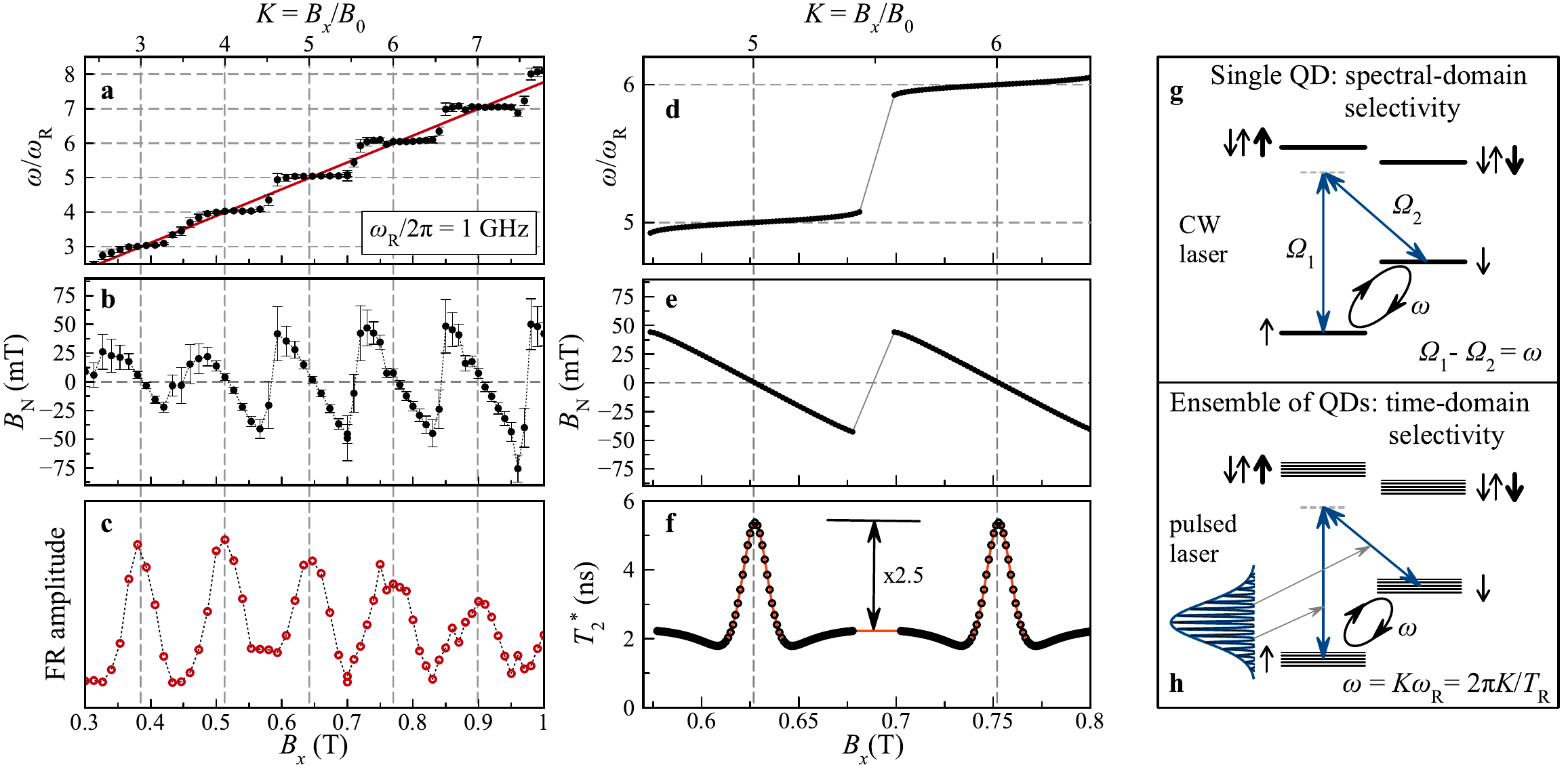}
		\caption{\textbf{Discretization of precession frequency by nuclear polarization.} \textbf{a} Electron precession frequency in units of repetition frequency of the laser vs. external magnetic field in absolute units (lower axis) and in units of $B_0=128\,$mT (upper axis) - the black dots. Error bars give the frequency fitting error with Eq.~\eqref{eq1}. Red line shows the electron precession frequency dependence on the magnetic field without the contribution of nuclear spin polarization. \textbf{b} Overhauser-field dependence on external magnetic field. The dotted line is a guide to the eye. \textbf{c} Amplitude of the Faraday rotation signal measured vs. $B_x$. The data in panels \textbf{a}-\textbf{c} are measured for a negative pump-probe spectral detuning: $E_{\text{Pu}}=1.3839$\,eV and $E_{\text{Pr}}=1.3864$\,eV. \textbf{d} Modeling of the frequency dependence on the external magnetic field. Plateaus around the modes 5 and 6 are clearly observable. \textbf{e} Magnetic field dependence of Overhauser-field built up in the QDs for compensation of the external field. \textbf{f} Variation of the spin dephasing time with external magnetic field. The multiplication factor of 2.5 is expected based on the simulation parameters. \textbf{g} Schematics of the energy levels of a singly charged QD in Voigt geometry and the two lasers photon energies $\hbar \Omega_1$ and $\hbar \Omega_2$, slightly detuned from the trion transitions, driving the $\Lambda$ system for coherent population trapping. $\omega$ is the Larmor frequency. \textbf{h} Ensemble of singly charged QDs driven by a negatively detuned pulsed excitation in the Voigt geometry. The laser pulse can be represented as a frequency comb combining multiple CW frequencies that satisfy the CPT for QDs having different optical transition energies and $g$ factors. We highlight a pair of CW laser components by gray arrows for one contributing QD. An alternative explanation is also given in the main text.}
		\label{fig2}
	\end{center}
\end{figure*}

As the next step, due to the time-resolution limitations set by the electronics in the extended pump-probe scheme, we use the common pump-probe protocol and fit Eq.~\eqref{eq1} to the FR data taken for different external magnetic fields ($B_x$) for 1\,GHz excitation. $S_1$ is considered to be time independent here as $T_2^* \gg T_{\text{R}}=1$\,ns. 
The oscillation frequency should depend linearly on the external magnetic field, as shown in Fig.~\ref{fig2}a by the red line. The data of the Larmor frequency evaluated by Eq.~\eqref{eq1} are shown by the black dots in Fig.~\ref{fig2}a, and demonstrate a non-linear step-like dependence of $\omega$, normalized by the laser repetition rate $\omega_{\text{R}}$.

As one can see in Fig.~\ref{fig2}a, the electron spin precession frequency shows small deviations from the linear dependence in small magnetic fields ($B_x<0.5$\,T). Increasing the magnetic field leads to the appearance of pronounced plateaus in the frequency dependence. The positions of the plateaus are fixed by the PSC on integer numbers of full spin revolutions during $T_{\text{R}}$ or $\omega = K\omega_{\text{R}}$. The center of each plateau corresponds to $B_x = K B_0$ (the upper axis in Fig.~\ref{fig2}a). The origin of this dependence is related to a nuclear magnetic field ($B_\text{N}$) resulting from the mean value of the Overhauser field exerted by the dynamically polarized nuclear spins, acting on the electron spin in each dot. Depending on the external magnetic field, $B_\text{N}$ decreases or increases the total magnetic field seen by the electron. The subtraction of the linear dependence from the experimental data allows us to extract the amplitude of $B_\text{N}$ as function of $B_x$. One can see in Fig.~\ref{fig2}b that the maximal amplitude of $B_\text{N}$ reaches $50\,$mT and can be oriented parallel or anti-parallel to the external field.

Such plateaus in the dependence of the electron spin precession frequency on the external magnetic field were observed earlier for electron spins localized on Fluorine donors in ZnSe epilayers~\cite{Zhukov2018}. One can explain them in terms of a dynamic nuclear polarization in the following way: the non-resonant optical excitation of the trion resonance leads to the appearance of an effective magnetic field along the light propagation direction - the optical Stark field. This field is perpendicular to $B_x$, the electron spin precesses about the total magnetic field which is tilted relative to the $x$-axis. This leads to the appearance of a sizable component of electron spin polarization along the $x$-axis ($S_x$), which efficiently polarizes the nuclear spins along the external magnetic field $B_x$~\cite{Carter,Korenev2011}. The nuclear polarization plays the role of the additional field described in the previous paragraph -- the Overhauser field, which acts back on the electron spins~\cite{Abragam1961,Glazov_book}.

In the experiments presented in Fig.~\ref{fig2}, we use a negative optical detuning, where the energy of the probe at the trion resonance is higher than the pump excitation energy~\cite{Yugova2009,Carter}. For a negative optical detuning in combination with the negative sign of the electron $g$ factor in the (In,Ga)As QDs, the Overhauser field adds to the external $B_{x}$ for electron spins which do not satisfy the PSC, driving their frequency to the PSC-consistent value, i.e. a laser period-commensurate value. This leads to the plateau-like behavior seen in Fig.~\ref{fig2}a.
The Overhauser field $B_\text{N}$ reaches the maximal amplitude of about 50\,mT when the external field is slightly larger than $B_{x} = 0.5KB_0$ (see Fig.~\ref{fig2}b). Its amplitude decreases with increasing $B_x$ and becomes zero at $B_x = KB_0$, the center of the plateau. A further increase of $B_x$ changes the direction of $B_\text{N}$. Here, it reaches the maximal negative amplitude slightly below $B_x = 0.5KB_0$. 

Figure~\ref{fig2}c demonstrates the value of $S_1$ in Eq.~\eqref{eq1} determined from the fits to the data as function of $B_x$, demonstrating a strong modulation. The magnetic field positions of the peaks correspond to integer spin precession periods within $T_{\text{R}}$, i.\,e. to fulfilled PSC. This allows us to assume that the $T_2^*$ time should be similarly modulated, as the amplitude $S_0$ in Eq.~\eqref{eq1} is expected to stay constant across the plateau, due to the constant values of the Larmor frequencies. To understand this behaviour we use the theory presented in Ref.~\cite{Korenev2011}, which relates the extension of the spin dephasing time at the plateau centers to the feedback strength between the electron and nuclear systems, and to the reduction of the nuclear spin fluctuations (the variance of the Overhauser field) (see Supplementary Note 4 and Supplementary Note 5 for more details). 

Figure~\ref{fig2}d demonstrates the simulation of the frequency behavior (normalized by the laser repetition frequency $\omega_{\text{R}}$) as a function of the external magnetic field $B_x$. The bottom scale gives the applied field, while the top one is normalized by the mode separation $B_0$. One finds fully developed plateaus around the modes 5 and 6. Figure~\ref{fig2}e shows the Overhauser field $B_{\text{N}}$ building up in the QD system as function of$B_x$. The parameters of the modeling are given in the Supplementary Note 5.

As suggested in Ref.~\cite{Zhukov2018}, for the electron spins satisfying the PSC the strong feedback should lead to a reduction of the nuclear spin fluctuations and, as a result, the spin dephasing time of the ensemble $T_2^*$ should be prolonged. As soon as $B_x$ differs from $B_x = KB_0$, the nuclear fluctuations recover due to the reduced feedback strength (see Supplementary Note 5). The dynamical nuclear polarization process looses its efficiency, even though the $x$-component of the electron spin polarization is largest for $B_x = 0.5KB_0$. At this field ($B_x = 0.5KB_0$), the amplitude of $B_\text{N}$ becomes redirected within a relatively narrow magnetic field interval. 

The magnetic field variation of the spin dephasing time $T_2^*$ calculated by Eqs.~(1)-(6) in the Supplementary Note 5, is demonstrated in Fig.~\ref{fig2}f. Depending on the magnetic field, this time becomes strongly modulated due to the periodic changes of the amplitude of the nuclear fluctuations. For the parameters used in our modeling, we expect a prolongation of the $T_2^*$ time by a factor of 2.5. 

The process of reduction of the nuclear field fluctuations at the center of the plateaus without build-up of nuclear polarization can be qualitatively understood in a similar way as the process of coherent population trapping~(CPT) suggested for a single QD~\cite{StepanenkoPRL06,Xu2008,XuNature09,Korenev2011,EthierPRL17}. As a reminder, once the difference of the photon energies of two linearly polarized continuous wave (CW) lasers $\Omega_1$ and $\Omega_2$ is equal to the Zeeman splitting of the ground state electron spin ($\uparrow$ and $\downarrow$), the system goes into a coherent dark state without the possibility of photon scattering into the excited trion state, see Fig.~\ref{fig2}g. Due to the nuclear spin fluctuations, the electron Zeeman splitting varies, moving the system out of the dark state. This leads to enhanced driving of one of the two optical transitions that causes scattering of photons and pulls the Zeeman splitting back to that of the dark state by changing the nuclear spin orientation in the surrounding. Such locking into the dark state induces the reduced variance of the Overhauser field.

In the case of pulsed excitation, as we use in our demonstration, the situation of reduction of the nuclear fluctuations can be seen in a similar way. However, in this case a short laser pulse can be presented as a combination of many CW lasers (frequency comb) with different frequencies. Therefore, there is a set of different two-frequency-combinations ($\Omega_1$ and $\Omega_2$) separated by different electron Larmor frequencies~\cite{Korenev2011}. These combinations can satisfy the CPT conditions for different QDs with corresponding spread of trion transitions, present in an inhomogeneous ensemble of QDs, see Fig.~\ref{fig2}h. Furthermore, one can represent the process of periodical pulsed excitation in an alternative way. In the transverse magnetic field, the pulsed circular excitation leads to creation of a coherent superposition of the ground state spin states (shown by the multiple lines for the ensemble). This superposition precesses in the magnetic field at the Larmor frequency $\omega=g\mu_{\rm B} B_x/\hbar$. When this frequency is commensurate to the laser repetition frequency $\omega_{\text{R}}=2\pi K /T_{\text{R}}$, the efficiency of spin polarization is strongly enhanced~\cite{Greilich2006}. Once the electron spin oscillates at one of these frequencies, it can also be seen as locked in a coherent dark state, as the Pauli principle forbids further excitation of the spin by circularly polarized pulses. If the nuclear field fluctuations bring the Zeeman splitting (or the Larmor frequency) out of the resonance condition, the interaction with the nuclear surrounding pulls the frequency back to the dark state, leading similarly to a reduction of the variance of the Overhauser field~\cite{Greilich2007,Carter}. In comparison to CW lasers, the pulsed excitation allows us to excite a spectrally broad distribution of QD transitions and can be seen as universal tool without strict requirement concerning the excitation laser energies for spectrally different QDs.

\subsection*{Two-laser protocol}
The relaxation dynamics of the contributing electron and nuclear spins differ by several orders of magnitude. The spin lifetime ($T_1$) of the resident electrons in the studied QDs was previously measured, reaching 1.7\,$\upmu$s~\cite{ZhukovPRB18}, while the lifetime of the nuclear spins for this sample ranges from several seconds under laser illumination up to hours in darkness~\cite{Greilich2007}. We want to make use of this difference and implement a protocol that suppresses the nuclear fluctuations by the 1\,GHz excitation and subsequently allows us to manipulate the electron spins with an arbitrary laser source in the prepared nuclear environment. Supplementary Note 2 shows such an implementation of the suggested alternating driving by the two available lasers. At this point we note, that in our experiment, the reduction of nuclear spin fluctuations is still present at the timescale of seconds without driving the system by the GHz laser. This observation seems to contradict previous experiments in single QDs, where the prepared state with reduced nuclear fluctuations decays with 46\,ms~\cite{EthierPRL17}. However, as measurements of the same group demonstrate, for the nuclear system with reduced nuclear fluctuations (variance) and increased nuclear polarization (mean value of the Overhauser field) the relaxation becomes biexponential with longer times reaching seconds~\cite{Gangloff2020}. It demonstrates that the difference in the relaxation times might depend on how the nuclear system is prepared, which requires further investigations.\\ As the measurement with alternating lasers takes a long time for the experiments (about 3 hours for one temporal trace), we present here an alternative realization of this idea.

\begin{figure}[t]
	\begin{center}
		\includegraphics[width=\columnwidth]{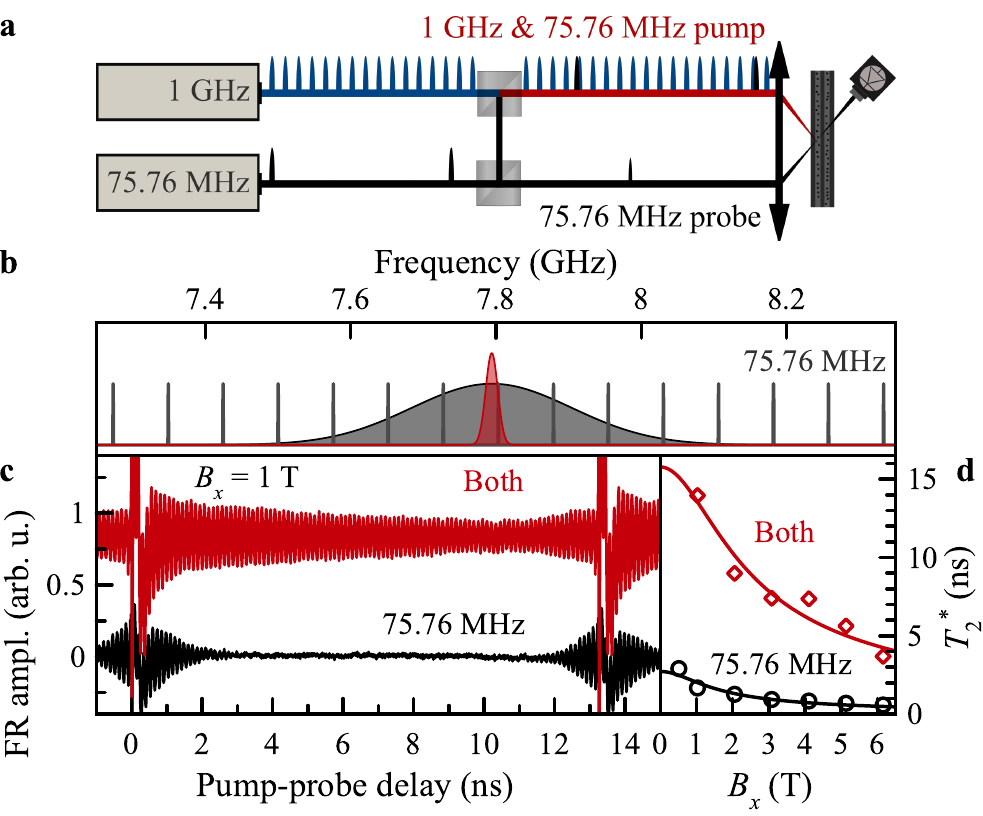}
		\caption{\textbf{Two-lasers protocol.} \textbf{a} Schematics of the protocol with two lasers applied to the same ensemble of QDs. The pump pulses of both lasers are applied simultaneously, while only the 75.76\,MHz laser probes the sample. \textbf{b} PSC modes for the 75.76\,MHz excitation with the frequency distributions in the unexcited ensemble of electrons, covering 7 modes for $B_x=1$\,T (black) and for the 1\,GHz excited ensemble, covering only a single PSC mode (red). \textbf{c} Pump-probe traces for the 75.76\,MHz case (black) and for both lasers applied (red) at $B_x=1$\,T. \textbf{d} The spin dephasing time $T_2^*$ determined for the case without the 1\,GHz laser illumination (black) and with the 1\,GHz laser (red) vs. magnetic field. Lines are fits using Eq.~\eqref{eqT2star}. All lasers are degenerate at $E_{\text{Pu/Pr}}=1.3867$\,eV.}
		\label{fig3}
	\end{center}
\end{figure}

Figure~\ref{fig3}a demonstrates a scheme, in which both lasers (75.76\,MHz and 1\,GHz) are applied simultaneously to the same QD ensemble. In this case we measure a pump-probe trace using the 75.76\,MHz laser while pump pulses of the 1\,GHz laser simultaneously excite the same ensemble, without any synchronisation between the lasers. Figure~\ref{fig3}c demonstrates a comparison between the case when only the 75.76\,MHz laser is applied (black) and the situation with both lasers (red). As one can see, in the latter case the dephasing of the ensemble is strongly reduced, which is a direct demonstration of a strong reduction of the frequency spread compared with the pure 75.76\,MHz case. As the emissions of the two lasers are not synchronized to each other, the Faraday rotation measured by the probe pulses only stems from the electron spins oriented by the pump pulses with the repetition period of 75.76\,MHz. There is still some minor mode-locking signal, as seen by the weak signal increase before the pump at 13.2\,ns, which might arise from the QDs not excited by the GHz laser. The sketch in Fig.~\ref{fig3}b demonstrates the calculated mode distributions for both cases, with the corresponding changes in the frequency distributions of the ensemble given by the spreads $\Delta g$ and $\delta B_{\text{N}}$, the colors correspond to the traces in Fig.~\ref{fig3}c.

Using the two-laser approach we measure the magnetic field dependence of $T_2^*$ for the two cases, using only the 75.76\,MHz laser and using both lasers applied. Figure~\ref{fig3}d demonstrates such a measurement, where the pump-probe traces are measured at the center of plateaus for a set of magnetic fields. For the analysis of these traces we took into account that the signal of a single mode oscillation for the two-laser approach can interfere with the multi-mode signal from the 75.76~\,MHz laser applied alone (see Supplementary Note 3 for a detailed trace analysis).

We characterize the dephasing behaviour of the extracted signal using the form:
\begin{equation}
T_2^*=\hbar/\left[\mu_B \sqrt{(\Delta g B)^2+(g \delta B_\text{N})^2}\right].
\label{eqT2star}
\end{equation}
This leads to the following fit values: (i) 75.76\,MHz only, $\Delta g = 4\times 10^{-3}$, $\delta B_\text{N}= 7.5$\,mT and (ii) both lasers, $\Delta g = 4.2\times 10^{-4}$, $\delta B_\text{N}= 1.3$\,mT. The reduction of the $g$-factor dispersion by one order of magnitude can be explained by the reduction of the frequency spread to a single mode, which is also additionally reduced in width by the NIFF. The reduction of the nuclear fluctuations $\delta B_\text{N}$ for the whole QD ensemble can be extracted from the width extrapolated to $B_x=0$ and gives a factor of 5.8, which is comparable to the value of 12 achieved in optimal conditions for a single QD, using the coherent population trapping technique~\cite{EthierPRL17}. Note that our experiment demonstrates a higher reduction than the factor 2.5 suggested by our model calculation, see Fig.~\ref{fig2}f. Taking into account the simplicity of the model, based purely on the optical Stark effect~\cite{Cohen-Tannoudji1972,Zhukov2018} and collinear hyperfine interaction~\cite{Urbaszek2013}, it still gives a good estimate. Another approach, using the collinear hyperfine interaction, is presented by Ref.~\cite{OnurPRB18}, demonstrating optimal reduction of the dephasing time $T_2^*$ by a factor of 6.75 for the Si-doped GaAs system. Furthermore, considering the inherent strain environment of our QDs, the noncollinear type of interaction mediated by the quadrupolar moments of the nuclei is expected to play an important role and may increase the influence on $T_2^*$, see Refs.~\cite{LattaNatPhys09,Hoegele2012,EthierPRL17,GangloffScience19}.

\section*{Discussion}

The $1\,$GHz laser repetition frequency used in this study allows us to explore the electron-nuclear spin dynamics for a pure single-mode Larmor spin precession in the inhomogeneous ensemble of (In,Ga)As/GaAs QDs. This is the first experimental realization of such a situation, which allows us to demonstrate the discretization of the total magnetic field acting on the electron spins. Furthermore, we confirm that at the center of the frequency plateaus, the nuclear spin fluctuations become reduced without build-up of a dynamic nuclear polarization, a situation comparable to the coherent population trapping experiments performed on single quantum dots. The pulsed excitation relaxes the requirement of a strictly accurate spectral tuning of the lasers (as required for a single QD) and makes this technique more universal. Additionally, the suggested two-laser protocol opens up a promising way to establish a reduced nuclear spin fluctuation surrounding using a high repetiton laser oscillator, while the lower repetition laser can be used for readout and manipulation of a large ensemble of spins.

\section*{Methods}
\subsection*{Sample}
The (In,Ga)As/GaAs QD ensemble was grown by molecular beam epitaxy on a (100)-oriented GaAs substrate. Adjacent sheets in the 20 QD layers are separated by $80\,$nm wide GaAs barriers. Resident electrons are provided by a $\updelta$ doping layer of Silicon placed $16\,$nm above each layer. The sample is thermally annealed at a temperature of $945\,^{\circ}$C for 30\,seconds to homogenize the QD size distribution and to shift the average transition energy to $1.39\,$eV.\\

\subsection*{Setup}
The electron spin polarization is measured at a sample temperature of $T=5.3\,$K using pump-probe spectroscopy in an external magnetic field $B_x$ applied perpendicular to the light propagation (Voigt geometry). Two lasers are used. The first one is a Ti:Sapphire laser with a pulse duration of $2\,$ps, a spectral full width at half maximum (FWHM) of $0.9\,$meV, and a pulse repetition frequency of $75.76\,$MHz (repetition period of $13.2\,$ns). The second laser is a Ti:Sapphire laser with a pulse duration of $150\,$fs and a repetition frequency of $1\,$GHz (repetition period of $1\,$ns). The pulses of the 1\,GHz laser are spectrally shaped using two sets of holographic gratings and slits (one set for pump and one for probe) to reach about $0.9\,$meV FWHM (duration of 1.5\,ps). The gratings enable us to introduce an energy detuning between the pump and probe beams. Both lasers are not synchronized or phase-locked.

The experiments presented in Fig.~\ref{fig1} and Fig.~\ref{fig3} are carried out using the degenerate case of pump-probe energies and external magnetic fields fixed at the integer precession modes, $B_x = KB_0$. At these magnetic fields, the overall behaviour is determined only by the negatively detuned electron transitions and it makes no difference if one uses degenerate or negatively detuned pulses. We found it experimentally easier to implement the degenerate case whenever possible as it requires one parameter less to control. For the experiments presented in Fig.~\ref{fig2}, the energy detuning plays an important role as it determines the direction of the Overhauser field while the external magnetic field is varied.

The time-resolved measurements are enabled by mechanical delay lines. To reduce the impact of scattered light, a double modulation scheme is used. The pump is helicity-modulated between left- and right-circular polarization using a photo-elastic modulator with a frequency of $84\,$kHz. The probe is intensity modulated with a frequency of $100\,$kHz while being vertically polarized. The signal is measured by a lock-in amplifier using the difference frequency of $16\,$kHz as a reference. The pump beams of both lasers are sent through the same lens and are focused to a spot diameter of $50\,\upmu$m. The probe beams are focused to $40\,\upmu$m spots. In this way, approximately $5 \times 10^5$ QDs are excited at the same time. The Faraday rotation of the probe beam is proportional to the electron spin projection along the light propagation direction and is measured using an optical bridge consisting of a Wollaston prism to separate the linear polarizations and Si-based balanced photo diodes.

For the extended version of the pump-probe experiment, we use the 1\,GHz laser, where the pump and probe pulses are picked by electro-optical modulators~(EOM), hitting the sample in bunches. The pump and probe bunches are separated by an electronically controlled delay. As the devices used here are not fast enough to have a high extinction ratio between neighboring pulses within a nanosecond, the rising and falling edges of the bunches have varying pulse amplitudes within about $6\,$ns. This time sets a limit on the time resolution of the extended pump-probe and makes it not usable for decay times shorter than 6\,ns. The pump and probe pulses stay synchronized to each other and the varying phase of the EOMs relative to the laser repetition frequency mainly add an additional exponential decay proportional to the falling edge of the EOMs used to select the pulse bunches. Here we use 130 pump pulses and six probe pulses for the corresponding bunches. This pump-probe sequence is repeated with a period of $516\,$ns.

\section*{Data Availability}
The data that support the findings of this study are available from the corresponding author upon reasonable request.

\section*{Acknowledgements}
We are grateful to V.L. Korenev for valuable discussions. We acknowledge the financial support by the Deutsche Forschungsgemeinschaft in the frame of the International Collaborative Research Center TRR 160 (Project A1) and the Russian Foundation for Basic Research (Grant No. 19-52-12059). I.A.Yu. acknowledges the support by Saint-Petersburg State University Research Grant No. 73031758. A.G. acknowledges support by the BMBF-project Q.Link.X (Contract No. 16KIS0857). The AFM figure was provided by Claudia Bock, Ruhr-Universität Bochum.

\section*{Author Contributions}
E.E. and A.G. conceived the experiment. E.E. and N.E.K. carried out the experiment and took the experimental data. E.E., N.E.K., and A.G. analyzed the experimental data. N.E.K. and I.A.Yu. conceived the theoretical model. D.R and A.D.W. prepared the sample. E.E., N.E.K., I.A.Yu., D.R.Ya., M.B., and A.G. wrote the manuscript.

\section*{Competing Interests}
The authors declare no competing interests.



\end{document}


	
	\makeatletter
	\renewcommand{\fnum@figure}{Supplementary~\figurename~\thefigure}
	\makeatother
	
	\title{Supplementary materials: Suppression of nuclear spin fluctuations in an InGaAs quantum dot ensemble by GHz-pulsed optical excitation}
	
	\author{E.~Evers}
	\email[Email: ]{eiko.evers@tu-dortmund.de}
	\affiliation{Experimentelle Physik 2, Technische Universit\"at Dortmund, 44221 Dortmund, Germany}
	
	\author{N.~E.~Kopteva}
	\affiliation{Experimentelle Physik 2, Technische Universit\"at Dortmund, 44221 Dortmund, Germany}
	
	\author{I.~A.~Yugova}
	\affiliation{Spin Optics Laboratory, St. Petersburg State University, 198504 St.\,Petersburg, Russia}
	
	\author{D.~R.~Yakovlev}
	\affiliation{Experimentelle Physik 2, Technische Universit\"at Dortmund, 44221 Dortmund, Germany}
	\affiliation{Ioffe Institute, Russian Academy of Sciences, 194021 St.\,Petersburg, Russia}
	
	\author{D.~Reuter}
	\altaffiliation[Present address: ]{Department Physik, Universität Paderborn, 33098 Paderborn, Germany}
	\affiliation{Angewandte Festkörperphysik, Ruhr-Universität Bochum, 44780 Bochum, Germany}
	
	\author{A.~D.~Wieck}
	\affiliation{Angewandte Festkörperphysik, Ruhr-Universität Bochum, 44780 Bochum, Germany}
	
	\author{M.~Bayer}
	\affiliation{Experimentelle Physik 2, Technische Universit\"at Dortmund, 44221 Dortmund, Germany}
	\affiliation{Ioffe Institute, Russian Academy of Sciences, 194021 St.\,Petersburg, Russia}
	
	\author{A.~Greilich}
	\affiliation{Experimentelle Physik 2, Technische Universit\"at Dortmund, 44221 Dortmund, Germany}
	
	\maketitle
	
	\section*{Supplementary Note 1: Extended pump-probe}\label{extPP}
	
	As discussed in the main text and in the Supplementary Note 5, the feedback strength of the electron-nuclear interaction is maximum at the center of the precession frequency plateaus, which should lead to reduced nuclear spin fluctuations and, therefore, to an extension of the spin dephasing time $T_2^*$. In contrast, in between the modes at the jumps between plateaus, the feedback strength is strongly reduced, which should lead to a strong influence of the fluctuating nuclear field and a reduced spin dephasing time. To support this observation we measured the $T_2^*$ using an extended pump-probe scheme~\cite{Belykh2016}. In this case we use only the 1\,GHz laser, where the pump and probe pulses are picked by electro-optical modulators and hit the sample as trains of pulses. The pump and probe trains are separated by an electronically-controlled delay.
	
	\begin{figure}[b]
		\begin{center}
			\includegraphics[width=\columnwidth]{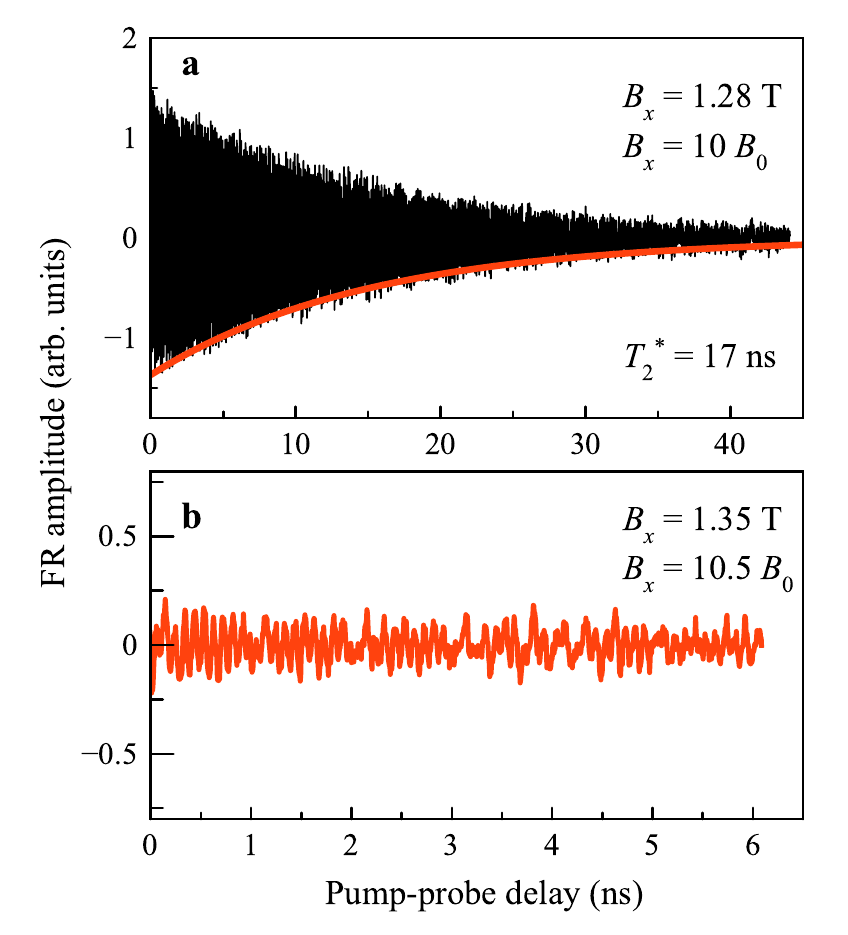}
			\caption{Long time dynamics of the electron spin polarization compared between the magnetic field on a mode \textbf{a} and in between modes \textbf{b}. On a mode, the electron spin polarization is amplified and decays within $T_2^*=17 \,$ns. In between modes the amplitude is tenfold lower and decays fast below the noise limit. Note that panels \textbf{a} and \textbf{b} cover different time scales. $E_{\text{Pu}}=1.3878$\,eV, $E_{\text{Pr}}=1.3864$\,eV.}
			\label{figS1}
		\end{center}
	\end{figure}
	
	The extended dynamics are shown in Supplementary Fig.~\ref{figS1} for an external field corresponding to $10\,B_0$ (on a mode, panel \textbf{a}) and to $10.5\,B_0$ (in between modes, panel \textbf{b}). Note the different time scales in panels \textbf{a} and \textbf{b}. On the mode, the electron spin polarization exhibits a $T_2^*$ time of $17\,$ns, which coincides also well with the time presented in the main text in Fig.~3d for the 1\,T case with both lasers applied. In between the modes, the amplitude is strongly reduced (about ten times) and is close to the noise level, which hints towards a much smaller $T_2^*$ time, as the accumulated spin amplitude depends directly on it.

	\section*{Supplementary Note 2: Alternative driving and readout}\label{alternative}
	
\begin{figure}[t]
		\begin{center}
			\includegraphics[width=\columnwidth]{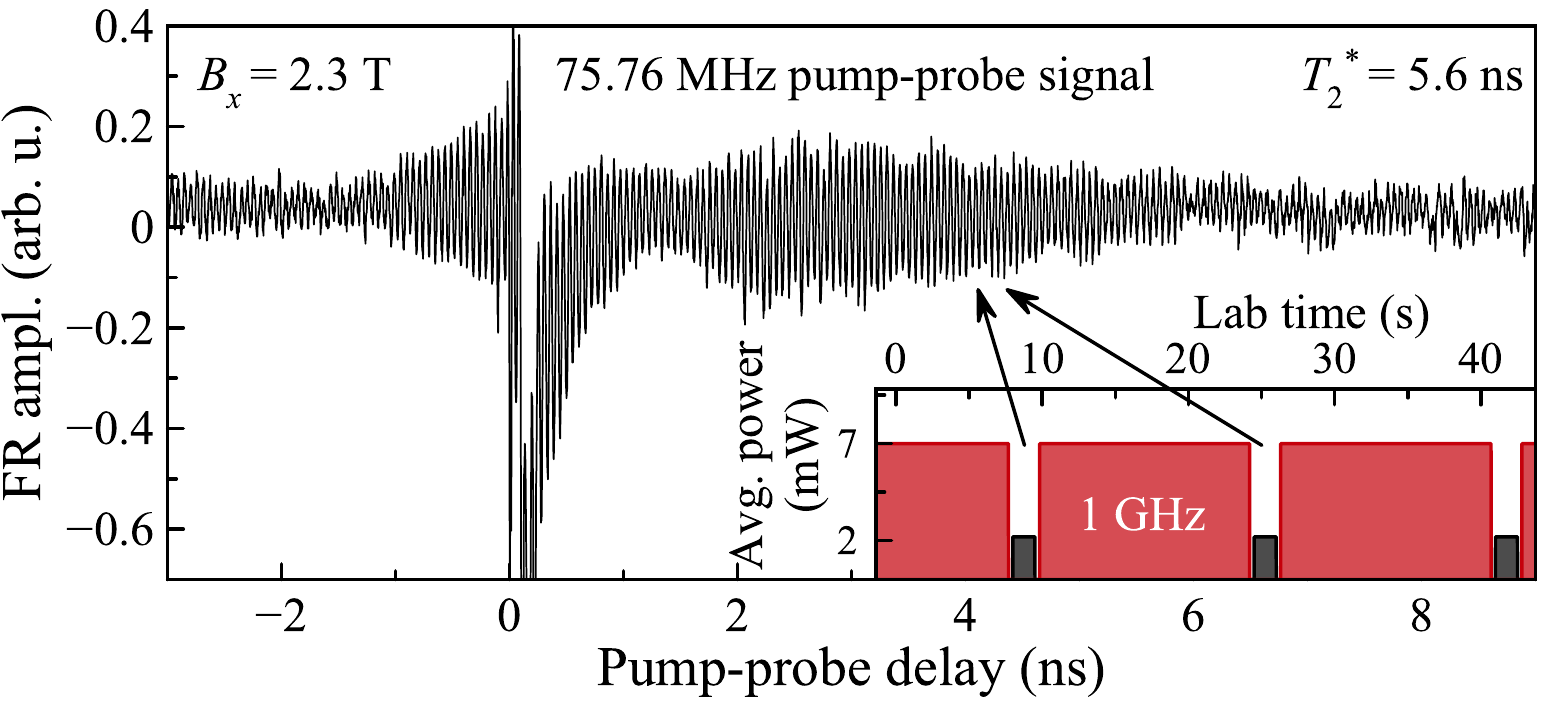}
			\caption{Alternating switching between the 1\,GHz and 75.76\,MHz lasers. The 1\,GHz laser is used to polarize the system at one mode, while the 75.76\,MHz laser is blocked. The measurement is done by the 75.76\,MHz laser, while the 1\,GHz laser is blocked. All lasers are degenerate at $E_{\text{Pu/Pr}}=1.3867$\,eV.}
			\label{figS2}
		\end{center}
	\end{figure}

	Supplementary Figure~\ref{figS2} shows the pump-probe signal for the following measurement: the magnetic field is fixed close to the center of a plateau at 2.3\,T.
	
	The 1\,GHz laser is switched on for 15\,s, then switched off and within 0.3\,s the sample is kept in darkness. Then the pump and probe from the 75.76\,MHz laser are switched
	on and the measurement for 10 steps (with 100\,ms of lock-in integration for each step) of the delay line is done, which takes about 1.5\,s. Then the laser is switched off, and after 0.3\,s of darkness the 1\,GHz laser is switched back on for 15\,s. This cycle is repeated for the full pump-probe picture. An elongated dephasing time ($T_2^*=5.6$\,ns) is observed. However, accumulation of this time trace took about 3 hours. Note, that the dephasing time is slightly shorter than with the 1\,GHz laser switched on all the time ($T_2^*=8$\,ns), but much longer than without it ($T_2^*=1.2$\,ns), see the main text for Fig.~3d. The timescale of the 1\,GHz-laser influence discovered in our experiment (several seconds) is puzzling if compared with the several milliseconds of fluctuation reduction timescales in the Ref.~\cite{EthierPRL17}. However, it is well supported by the Ref.~\cite{OnurPRB18}, where the variation of $T_2^*$ is happening on the timescale of nuclear spin diffusion, which is on the order of several seconds under laser illumination in our case. These statements require additional investigations. Making the measurement interval for the 75.76\,MHz longer or the polarization time for the 1\,GHz laser shorter resulted in a reduction of the signal. Such a measurement demonstrates the implementation of the possibility to use the 1\,GHz laser for a preparation of the nuclear system for a subsequent measurement with another laser system for about 1\,s.
	
	\section*{Supplementary Note 3: Decomposition of the time-traces for two-lasers protocol}\label{2lasers}
	
\begin{figure}[th!]
		\begin{center}
			\includegraphics[trim=0mm 0mm 0mm 0mm, clip, width=\columnwidth]{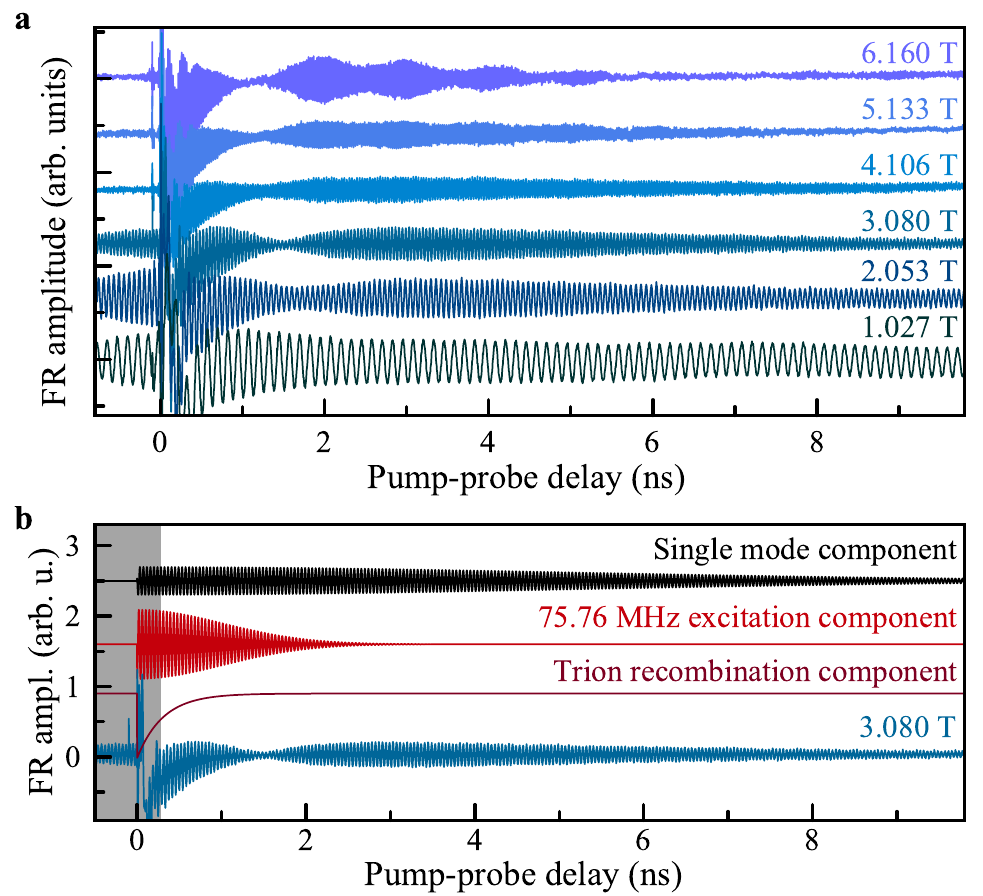}
			\caption{\textbf{a} Temporal dynamics measured for different magnetic fields corresponding to plateau centers. \textbf{b} Decomposition of the contributing components for the $B_x=3.08\,$T signal. Grey area is outside of the fitting area.}
			\label{figS3}
		\end{center}
	\end{figure}

	The Supplementary Fig.~\ref{figS3} demonstrates pump-probe traces measured at the plateau centers for different magnetic fields. The dephasing times of the extended single-mode time dynamics are extracted from the signals taking into account that both lasers can contribute to the signal, see the exemplary decomposition of the signal for $B_x=3.08$\,T in Supplementary Fig.~\ref{figS3}b. One can clearly separate three contributions: the non-oscillating trion decay, which disappears within $0.4$\,ns; the multi-mode component with a Gaussian decay originating from the quantum dots (QDs) only excited by the 75.76\,MHz laser (red); and the slowly decaying single mode component created by the common action of both lasers (black). The dephasing times for this last component are presented in the main text in Fig.~3d.
	
	Additionally, it is clearly seen that above $B_x=5$\,T an additional beating structure starts to appear, signaling that the spread of the frequencies for the 1\,GHz laser is increased above the mode separation, see Supplementary Note 4 for more details. 
	
	\section*{Supplementary Note 4: Single-mode and plateau-size limits}\label{limits}
	
	\begin{figure*}[th!]
		\begin{center}
			\includegraphics[trim=0mm 0mm 0mm 0mm, clip, width=2.0\columnwidth]{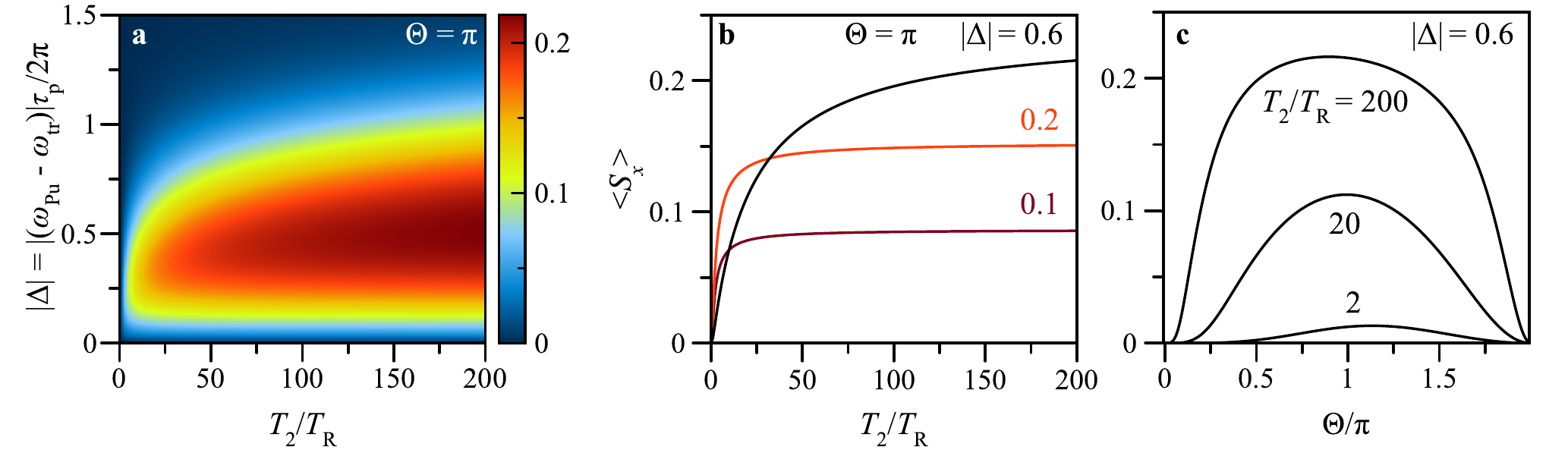}
			\caption{\textbf{a} Calculated color map of $\langle S_x \rangle$ as a function of $T_2/T_{\rm R}$ and the absolute value of optical detuning $|\Delta|$ between the energy of the trion resonance ($\hbar \omega_{\text{tr}}$) and the excitation energy ($\hbar \omega_{\text{Pu}}$) with the pulse duration $\tau_p$. The pulse area is $\Theta=\pi$. \textbf{b} Three slices of panel \textbf{a} are shown for different detunings to demonstrate the amplitude evolution and saturation at high values of $T_2/T_{\rm R}$. \textbf{c} Power dependence for three ratios of $T_2/T_{\rm R}$ at fixed detuning $|\Delta|=0.6$, demonstrating that weak excitation leads to a smaller electron spin polarization and therefore a weaker nuclear field.}
			\label{figS4}
		\end{center}
	\end{figure*}
	
	Here we discuss the factors limiting the single-mode condition and the maximal length of the plateaus for the QDs studied in this paper. At high repetition frequencies (small repetition periods) the limitation is given by the trion recombination time, $\tau_r$. As is known from Ref.~\cite{Greilich2006}, the degree of the electron spin polarization under pulsed excitation depends on the recombination of the trion state. For $T_{\rm R} < \tau_r$, the efficiency of the spin initialization becomes strongly reduced. Therefore, one can set the qualitative limit for $T_{\rm R}$ at $3\tau_{\text{r}}$ at which the trion decays by about 95\% on average. In the studied sample $\tau_{\text{r}}=0.4$\,ns, so that $3\tau_{\text{r}} = 1.2$\,ns. For $T_{\rm R}=1$\,ns we are close to this condition. This limitation can be reduced, by placing the QDs into an optical microcavity and using the Purcell effect to enhance the spontaneous emission~\cite{PurcellAPL03} and to shorten $\tau_{\text{r}}$.
	
	Furthermore, in the case of $\Delta g=0$, the limitation for the low repetition frequency side is set by the nuclear fluctuations $\delta B_{\text{N}}$. The mode distance in magnetic field is defined as $B_0 = \hbar \omega_{\text{R}} /(g \mu_{\rm B})$, with $\omega_{\text{R}}=2\pi/T_{\rm R}$. The nuclear field fluctuation is defined as the half-width at half-maximum. So, to be able to reach the single mode regime, one has to overcome at least $3 \delta B_{\text{N}}$ (a half of $B_0$). Therefore, $T_{\rm R}<2\pi\hbar/(g \mu_{\rm B} 3 \delta B_{\text{N}})=5.6$\,ns. It is useful to remind here, that the fluctuating nuclear field is proportional to $1/\sqrt{N}$, with $N$ being the number of the nuclear spins in the QD. This means that for bigger QDs, the fluctuating field is reduced and the limit set on $T_{\rm R}$ can be relaxed, a slower repetition laser can be used. An additional effect of the bigger QD size is the acceleration of the trion recombination ($\tau_r$ becomes smaller)~\cite{GreilichPRB06}, which is preferential for smaller $T_{\rm R}$. Bringing both limits together, for a robust single-mode regime with highest electron spin polarization, the repetition period of the laser should be in the range $1.2$\,ns\,$<T_{\rm R}<5.6$\,ns.
	
	In the previous paragraph we have assumed that $\Delta g=0$. If it is nonzero, this parameter sets an additional limitation on the applied external magnetic field, as the spread of frequencies is linearly increasing with field strength. For $\Delta g=0.004$ and $T_{\rm R}=1$\,ns, $M=2$ starts to become possible at $B_x>5.9$\,T, see Eqs.~(1) and~(2) of the main text. This conclusion is supported by the observation of a beating in the trace in Supplementary Fig.~\ref{figS3}a, recorded for $B_x=6.16$\,T. Note that once the number of possible modes $M$ is increased above 2, the dephasing is defined by $\Delta g$. This is the case for the 75.76\,MHz laser, presented by the black circles in Fig.~3d. However, if the mode separation is only allowing one mode, the spread $\Delta g$ is effectively reduced by the effect of NIFF. Furthermore, $\Delta g$ can be selected by the spectral width of the laser, as described in the experimental details, or by choosing another growth technique for the QDs. For example, the $\Delta g$ can be strongly reduced using QDs grown by infilling of droplet-etched nanoholes~\cite{LoeblPRB19} or many-electron GaAs/(Al,Ga)As QDs~\cite{MarkmannNatCom19}, which avoids the use of Indium atoms that affects the $g$-factor values strongly.
	
	Finally, for any given $T_{\rm R}$ the maximal length of the plateaus, limited from top by the distance between the modes $B_0 = 2 \pi \hbar /(g \mu_{\rm B} T_{\rm R})$, is determined by the maximal Overhauser field $B_{\text{N,max}}$, see Supplementary Fig.~\ref{figS4} and the following Supplementary Note 5. It is dependent on the types of constituent nuclear species, the leakage factor $f_{\text{N}}$, and the electron spin polarization $\langle S_x \rangle$ along the direction of $B_x$. The $\langle S_x \rangle$ in turn depends on the optical detuning between the pump and probe pulses $\Delta$, the optical pump power $\Theta$, the electron spin coherence time $T_{\text{2}}$, and $T_{\rm R}$~\cite{KoptevaPSSB19}, as demonstrated in Supplementary Fig.~\ref{figS4}.
	
	Supplementary Figure~\ref{figS4}a shows a color map calculated for the optical pulse power of $\Theta = \pi$ in dependence on the optical detuning and the ratio $T_2/T_{\rm R}$. For the calculations the spin coherence time was assumed homogeneous, so the resident electron spin lifetime is equal to the spin coherence time. In our case $T_{\rm R}=1$\,ns and the coherence time of the electron spins $T_2$ for these dots is about 3\,$\mu$s, which makes the relation of $T_2/T_{\rm R}=200$ a secure underestimation. The dependence of $\langle S_x \rangle$ for several detunings is demonstrated additionally in the Supplementary Fig.~\ref{figS4}b. The amplitude saturates at $T_2/T_{\rm R}$ close to 200 and the maximal amplitude is reached at a detuning of $|\Delta|=0.6$. These dependencies show that variation of the detuning has a strong impact on the value of $\langle S_x \rangle$, and therefore on the maximal Overhauser field $B_{\text{N,max}}$. So, the maximal value of $B_{\text{N,max}}$ will be reached at $|\Delta|=0.6$. Finally, the Supplementary Fig.~\ref{figS4}c demonstrates the power dependence of $\langle S_x \rangle$ for $|\Delta|=0.6$. The maximum of the nuclear polarization should be observed for $\Theta=\pi$. This panel also shows that the maximal $\langle S_x \rangle$ is strongly influenced by the spin coherence time $T_2$. For strongly reduced $T_2$ values, the nuclear field is decreasing and so should the plateau length.
	
	Furthermore, the strain inhomogeneity of the self-assembled QDs affects the carrier spin coherence due to the involved quadrupolar effects of the nuclei~\cite{Urbaszek2013}. The electron spin coherence time $T_2$, observed on a similar types of QDs, is strongly reduced at lower magnetic fields (below 3\,T) and becomes increased at magnetic fields above 3-4\,T~\cite{BechtoldNatPhys15,GangloffScience19,Press2010, Stockill2016}. As the observed feedback mechanism in our paper directly depends on the $T_2$ of the electron spins, we observe a limitation for lower magnetic fields. For the higher side of the fields, the spread of the $g$ factors limits the feedback efficiency. The latter one is not present for the case of a single QD or it can be relaxed by using lattice-matched GaAs/AlGaAs QDs, grown by \textit{in situ} nanohole etching and infilling~\cite{Atkinson2012,Chekhovich2017}.
	
	\section*{Supplementary Note 5: Simulation of the plateau behavior} \label{simulation}
	
	The feedback strength in the electron-nuclear spin system can be described by the parameter $\lambda$~\cite{Korenev2011}:
	\begin{equation} \label{lambda}
		\lambda = -\frac{T_{1\text{e}} + T_\text{d}}{T_{1\text{e}}T_\text{d}} \left(1 - \frac{A f_\text{N} Q_\text{av}}{\hbar} \frac{\delta \langle S_x \rangle}{\delta \omega_\text{N}}\right),
	\end{equation}
	where $T_{1\text{e}}$ is the nuclear spin polarization time via the electron spin polarization, $T_\text{d}$ is the nuclear spin-lattice relaxation time, $A$ is the average hyperfine constant, and $f_{\text{N}}$ is the leakage factor. $Q_\text{av} = \sum_{i=j}^54I_{j}(I_{j}+1)n_{\text{QD},j}/3$ is the factor dependent on the nuclear spin ($I_i$) averaged over all nuclear species with spin ($I_{j}$) and fraction $n_{\text{QD},j}$ in an elementary cell with 2 atoms. $Q_\text{av} = 20$ for (In,Ga)As QDs with 35\,\% of In concentration. $\omega_\text{N}$ is the Larmor precession frequency in the Overhauser field.
	
	One can use following equation for $T_{1\text{e}}$ from Ref.~\cite{Glazov_book}:
	\begin{equation} \label{T1e}
		T_{1\text{e}} = \left( \frac{\hbar N}{A}\right)^2 \frac{1+\omega_{\text{N}}^2\tau_{\text{C}}^2}{2F\tau_{\text{C}}}.
	\end{equation}
	$\tau_\text{C}$ is the correlation time in the electron-nuclear spin system. The factor $F$ represents the average fraction of time during which the dot is occupied. We assume that $F = 1$ for a resident electron. 
	
	One can write the second term of Eq.~(\ref{lambda}) analytically using the terms $L$ and $M$ introduced in Ref.~\cite{KoptevaPSSB19}:
	\begin{equation} \label{Sx}
		\frac{\delta \langle S_x \rangle}{\delta \omega_\text{N}} = S_\text{t} \frac{(1+LM)\cos(\omega_\text{N}T_\text{R}) - (L+M)}{\left[1+LM - (L+M)\cos(\omega_\text{N}T_\text{R})\right]^2},
	\end{equation}
	
	\begin{equation} \label{Sx0}
		S_\text{t} = \frac{T_2 (1-Q^2) K}{4} \left[1-\exp(-T_\text{R}/T_2)\right].
	\end{equation}
	The functions $K$ and $Q$ are also described in Ref.~\cite{KoptevaPSSB19}.
	
	According to the supplementary material of Ref.~\cite{Zhukov2018} the spin dephasing time is defined by the characteristic square of the nuclear spin fluctuations $\langle \delta I_\text{N}^2\rangle$:
	\begin{equation}\label{eq:NF3}
		\langle \delta I_\text{N}^2\rangle = \frac{I(I+1)N|\lambda_0|}{|\lambda|},
	\end{equation}  
	where $\lambda_0 = \lambda(\omega_\text{N} = 0)$ is given by the nuclear depolarization time via the electron spin. The spin dephasing time ($T_2^*$) can be estimated using the approach from Refs.~\cite{MerkulovEfrosRosen,Glazov_book}:
	\begin{eqnarray}\label{eq:NF5}
		T_\text{s} = \frac{\hbar N}{A\sqrt{\langle (\delta I_\text{N})^2\rangle}}.
	\end{eqnarray}
	
	The simulations in the main text are done using the equations given in Refs.~\cite{Zhukov2018,KoptevaPSSB19} with the following parameters: $\Delta=-0.55$, $N=5.5\times 10^5$, the electron spin coherence time $T_2=2$\,$\mu$s, $\tau_\text{C}=10$\,ns, $T_\text{d}=2$\,min, $\Theta=\pi$, the electron $g$ factor $g=-0.57$, $Q_\text{av}=20$, $f_{\text{N}0}=0.86$, and $A = 49.2$\,$\upmu$eV.
	
